# LANL Report LA-UR-02-5837 (2002)

# Systematics of Proton-Induced Fission Cross Sections for Intermediate Energy Applications

A. V. Prokofiev[*]

*The Svedberg Laboratory, Uppsala University, Box 533, S-751 21 Uppsala, Sweden*

*Department of Neutron Research, Ångström Laboratory, Uppsala University, Box 525, S-751 20 Uppsala, Sweden*

*V.G. Khlopin Radium Institute, 2oi Murinskiy Prospect 28, Saint-Petersburg 194021, Russia*

and

S. G. Mashnik and W. B. Wilson

*Nuclear Physics Group (T-16), Theoretical Division, Los Alamos National Laboratory, Los Alamos, New Mexico 87545*

**Abstract** - The recent systematics of proton-induced fission cross sections is extended to a wider range of target nuclei and incident energies. Reasonable agreement with available experimental data is demonstrated. The extended systematics is employed to generate a data library for use in the CINDER'90 transmutation inventory code.

## I. CINDER'90 FISSION DATA NEEDS

The CINDER'90 code[1] is intended for tracking transmutation of nuclides in material. The calculations are deterministic and based on analytic expressions describing the transmutation process. The input data for the code include the initial composition of the material, a user-provided description of the particle flux, and data libraries for nuclear reaction cross-sections and radioactive decay. Destruction and production of nuclides beyond the particle and energy limits of the CINDER'90 library must be tallied and supplied as constant loss or gain terms during the irradiation history.

The CINDER'90 code is frequently used in connection with a Monte-Carlo transport code (*e.g.*, LAHET[2], MCNPX[3]), which provides the particle flux within the material, as well as the probabilities of production/destruction of nuclides, using nuclear reaction models. Statistical uncertainties of these probabilities are determined by the number of particles tracked. Minor reaction paths may be poorly sampled, even following a large number of particles at a great expense of CPU time. This is the case, in particular, for fission reactions resulting in a wide variety of products. In order to facilitate the use of the CINDER'90 code, calculations with nuclear reaction models have to be replaced by data evaluations and systematics.

The goal of the present work is to develop and validate systematics for proton-induced fission cross sections. The systematics must be capable of giving predictions for reactions involving stable and unstable target nuclei, because the latter may give important contributions to the transmutation process.

---

[*] Corresponding author, Tel. +46-18-471-3850, Fax +46-18-471-3833, E-mail: Alexander.Prokofiev@tsl.uu.se

II. FORMULATION OF THE SYSTEMATICS

The recent work of Prokofiev[4] includes parameterizations of the (p,f) cross sections versus the fissility parameter $x=Z^2/A$, where $Z$ and $A$ are the charge and the mass of the compound nucleus, correspondingly. The systematics[4] allows one to predict the (p,f) cross sections, in particular, for the following fissioning systems:

- from $^{197}$Au+p to $^{209}$Bi+p ($32.32 \leq x \leq 33.60$) in the energy region from 70 MeV to 30 GeV,
- $^{232}$Th+p and systems with higher fissility parameter ($x \geq 35.54$) in the energy region from 20 MeV to 30 GeV.

In the present work, the systematics of Ref. 4 has been extended to the following fissioning systems and energy range:
- from $^{197}$Au+p to $^{209}$Bi+p in the energy region from 35 to 70 MeV (Subsection IIA),
- between $^{209}$Bi+p and $^{232}$Th+p ($33.60 \leq x \leq 35.54$) in the energy region from 70 MeV to 30 GeV (Subsection IIB).

*II.A. Extension of the Systematics for Fissioning Systems between $^{197}$Au+p and $^{209}$Bi+p*

The (p,f) cross sections were measured for a few nuclei in the Au-Bi region at 35-70 MeV by Khodai-Joopari[5], Ignatyuk et al[6], and Gadioli et al[7]. The measured fission cross sections decrease rapidly as the incident energy decreases and approaches the fission barrier, which amounts to about 20-25 MeV for the studied fissioning systems[6]. For example, the $^{197}$Au(p,f) cross section at 35 MeV amounts to only about 0.1% of the total reaction cross section. For this reason, together with the paucity of the experimental database below 35 MeV, we have refrained ourselves from extension of the systematics to lower energies and lighter fissioning systems.

Prior to building parameterizations, we reviewed normalization procedures for the experimental data of Ref. 5-7. The measurements of Refs. 5 and 6 are relative, and the normalization of the published experimental data can be traced back to an earlier work of Huizenga et al[8]. The data of the latter work are, in turn, not sufficiently reliable because of the use of obsolete and poorly documented experimental techniques. The method of data normalization used in Ref. 7 is unknown and considered as insufficiently reliable, either. On the contrary, measurements of Shigaev et al[9] in the 70-200 MeV energy region were made absolute using well-documented determination of the proton beam intensity. Therefore, the results from Ref. 9 were employed to deduce re-normalization factors that amount to 0.496±0.049 for the data of Refs. 5 and 6, and 0.453±0.045 for the data of Ref. 7.

On the basis of the re-normalized experimental data sets of Refs. 5-7, the following approximation is suggested:

$$\sigma(E) = \sigma_0 \exp\left[-\frac{(E-E_0)^2}{2w^2}\right], \qquad (1)$$

where $\sigma$ is the (p,f) cross section (mb), $E$ is the incident energy (MeV), $E_0 = 76.3$ MeV, and $w$ and $\sigma_0$ are parameters that depend on the fissioning system and characterize, respectively, the steepness and the absolute scale of the fission excitation function.

As shown in Ref. 4, the parameters in the cross-section systematics above 70 MeV vary smoothly with the fissility parameter. This finding can be interpreted as a consequence of the "wash-out" of nuclear structure effects with growing excitation energy of the residual nuclei, which may undergo fission. The fissioning nucleus with sufficiently high excitation energy behaves like a structureless charged liquid drop. Therefore the probability of fission is governed by the fissility parameter that relates the Coulomb and surface energy of the nucleus.

At lower incident energies, structural effects manifest themselves in the fission excitation functions. In particular, as shown in Ref. 6, the shape of the energy dependence of the fission probability is different for nuclei that are spherical or deformed in the ground state, due to differences in collective enhancement of level density in the neutron emission channel.

Indeed, the parameters $w$ and $\sigma_0$ in Eq. (1) do not seem to depend regularly on the fissility parameter only, but include a component that correlates with the shell correction $\delta W_{gs}$ to the ground-state mass of the fissioning nucleus. Therefore the following systematics is suggested:

$$w(Z,A) = a + b\ (Z^2/A) + c\ \delta W_{gs}(Z,A), \qquad (2)$$

where the shell correction $\delta W_{gs}$ is calculated for the compound nucleus using the systematics of Myers and Swiatecki[10], and the parameters found with the least-squares method are $a$=-33.667, $b$=1.5699, and $c$=0.30069. Finally, the absolute scale parameter $\sigma_0$ in Eq. (1) is defined by requirement of the continuity of the excitation function at the boundary of the present systematics and the one of Ref. 4:

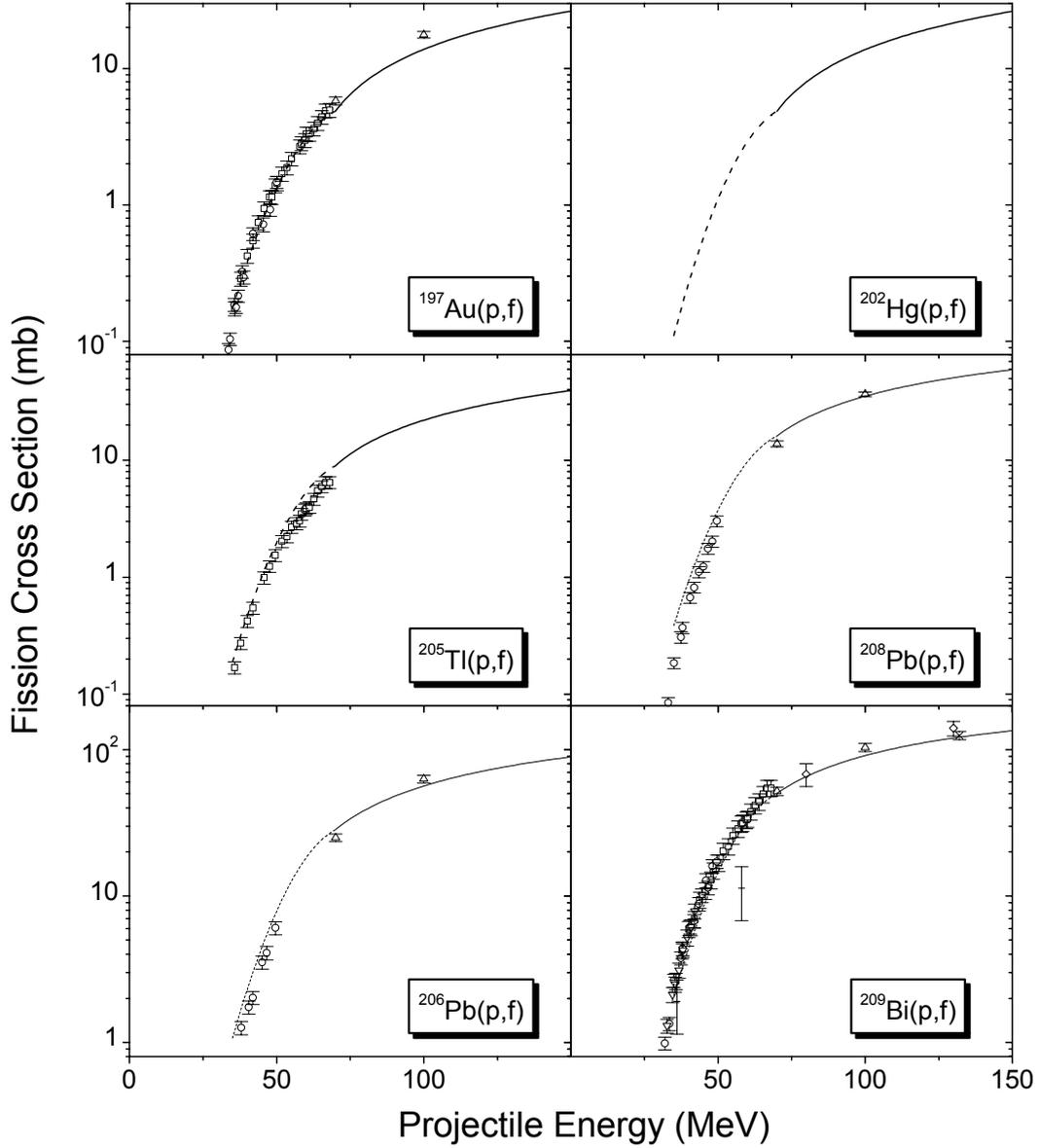

Fig. 1. Proton-induced fission cross sections of $^{197}$Au, $^{202}$Hg, $^{205}$Tl, $^{208}$Pb, $^{206}$Pb, and $^{209}$Bi nuclei versus incident proton energy. The symbols represent experimental data of Khodai-Joopari[5] (circles), Ignatyuk et al[6] (squares), Gadioli et al[7] (diamonds), Shigaev et al[9] (triangles), Beljaikin et al[11] (straight crosses), Sugihara et al[12] (skewed crosses), and Steiner et al[13] (stars). The solid lines show predictions of the systematics of Prokofiev[4]. The dashed lines represent extension of the systematics developed in the present work.

$$\sigma_0 = \sigma_b \exp\left[\frac{(E_b - E_0)^2}{2w^2}\right], \qquad (3)$$

where $E_b$ = 70 MeV, and $\sigma_b = \sigma(E_b)$ is calculated according to the high-energy systematics[4].

The predictions of both high-energy and low-energy systematics for proton-induced fission cross sections of $^{197}$Au, $^{205}$Tl, $^{208}$Pb, $^{206}$Pb, and $^{209}$Bi nuclei are shown in Figure 1 together with the experimental data[5-7, 9, 11-13]. In addition, the (p,f) cross section of $^{202}$Hg (the main constituent of natural mercury) is shown as an example of systematics predictions for an unmeasured excitation function.

Overall good agreement is observed between the systematics predictions and the experimental data. The root-mean-square difference between the systematics and the data amounts to 9% for $^{197}$Au, 17% for $^{209}$Bi, and about a factor of 2 for Pb isotopes. Taking into account that, in the latter case, only single data sets with uncertain normalization are available, the agreement seems satisfactory. In all cases, the low-energy and high-energy systematics have a smooth connection; the additional uncertainty of the cross section rising from unsmoothness of the connection does not exceed 7%.

*II.B. Extension of the Systematics for Fissioning Systems between $^{209}$Bi+p and $^{232}$Th+p*

The systematics of the earlier work[4] does not include provisions for fissioning systems between $^{209}$Bi+p and $^{232}$Th+p ($33.60 \leq x \leq 35.54$). Not a single data point exists at this domain, and therefore it is not possible to verify predictions of any systematics or model calculation. On the other hand, as shown in Ref. 4, the following cross section representation is valid for incident energies above 70 MeV for fissioning systems at both edges of this range:

$$\sigma(E) = P_1\{1 - \exp[-P_3(E - P_2)]\}(1 - P_4 \ln E), \qquad (4)$$

where $P_i$ (i=1…4) are fitting parameters. As seen in Figure 2, they depend smoothly on the fissility parameter.

As soon as no influence of nuclear structure effects is expected for the considered energy region above 70 MeV (see the previous subsection), it is suggested that the parameter values in the range $33.60 \leq x \leq 35.54$ are to be found by interpolation of the systematics predictions. On the basis of the data shown in Fig. 2, the logarithmic interpolation scheme is chosen:

$$\ln P_i = Q_{i1} + Q_{i2} x, \qquad (5)$$

where the constants $Q_{ij}$ ($i = 1…4, j = 1,2$) are calculated as follows:

$$Q_{i1} = \frac{x_{Th} \ln P_i(x_{Bi}) - x_{Bi} \ln P_i(x_{Th})}{x_{Th} - x_{Bi}}, \qquad (6)$$

$$Q_{i2} = \frac{\ln P_i(x_{Th}) - \ln P_i(x_{Bi})}{x_{Th} - x_{Bi}}, \qquad (7)$$

where $P_i(x)$ are predictions of the systematics of Ref. 4, and indexes "Bi" and "Th" denote the $^{209}$Bi+p and $^{232}$Th+p fissioning systems, correspondingly, for which abundant experimental data are available. The resulting $Q_{ij}$ values are:

$$Q_{11} = -27.74, \quad Q_{12} = 0.9906, \quad Q_{21} = 25.83, \quad Q_{22} = -0.6567,$$
$$Q_{31} = -45.80, \quad Q_{32} = 1.227, \quad Q_{41} = -10.95, \quad Q_{42} = 0.2320.$$

The resulting dependence of $P_i$ ($i = 1…4$) factors on the fissility parameter is shown in Fig. 2 together with the data from Ref. 4.

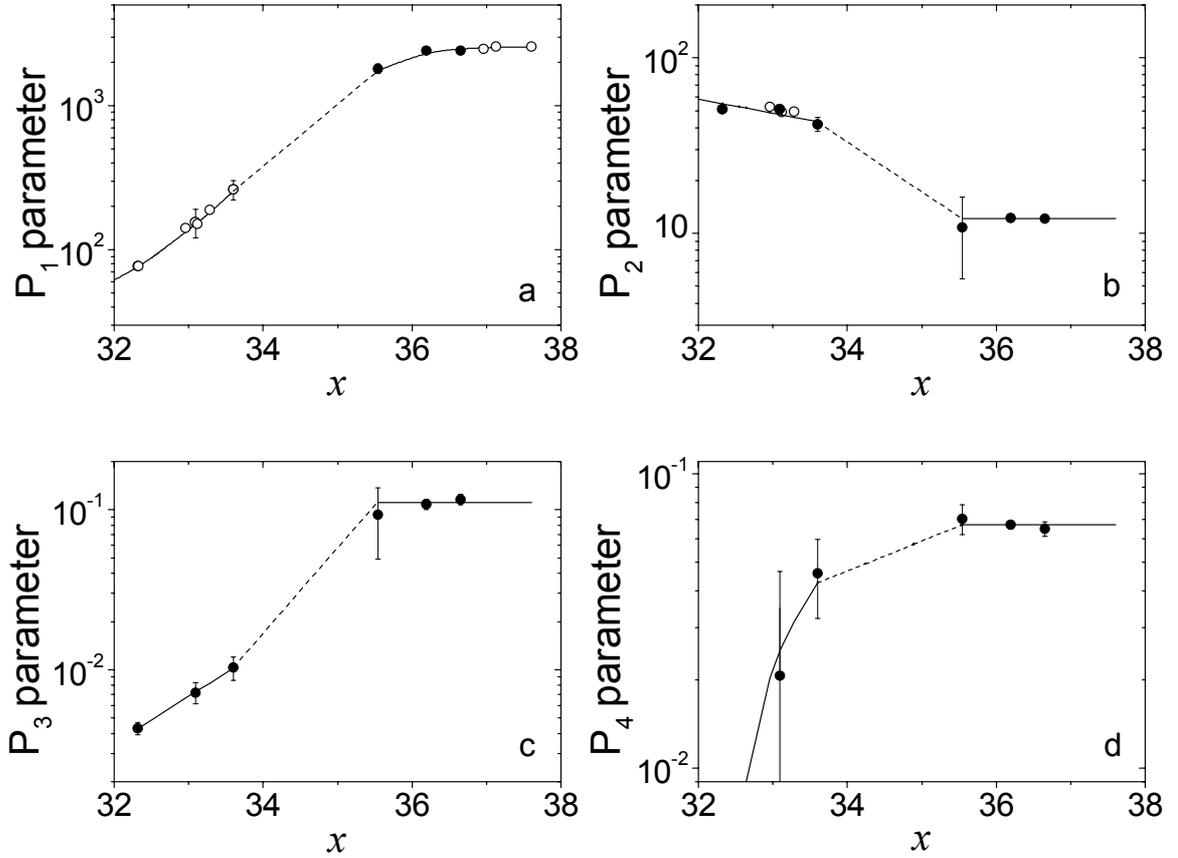

Fig. 2. Parameters $P_i$ ($i = 1…4$) in the (p,f) cross section systematics versus the fissility parameter $x = Z^2/A$ for the compound nucleus. The symbols and solid lines represent individual best fits and systematics predictions obtained by Prokofiev[4]. The dashed lines show extension of the systematics developed in the present work for fissioning systems in the range $33.60 \leq x \leq 35.54$.

As an example of application of the extended systematics, (p,f) cross sections of a few long-lived target nuclides ($^{210}$Po, $^{211}$At, $^{227}$Ac) are calculated and shown in Figure 3, in comparison with the predictions for $^{209}$Bi and $^{232}$Th provided by the systematics of Ref. 4.

The following features of the predicted fission excitation functions are seen in Fig. 3:

- an increase of the fissility parameter leads to higher cross sections for any incident proton energy in the considered region (70 MeV – 30 GeV),
- the maximum of the excitation function migrates from hundreds MeV to tens MeV as the fissility parameter increases,
- the slope of the excitation function after passing the maximum becomes steeper for fissioning systems with higher fissility.

### III. CONCLUSIONS AND FURTHER WORK

The systematics of proton-induced fission cross sections is extended to a wider range of target nuclei and incident energies. The extended systematics is employed to generate a data library for use in the CINDER'90 transmutation inventory code. Further work will include a similar effort for neutron-induced fission, and combination of our results with systematics for fission yields.

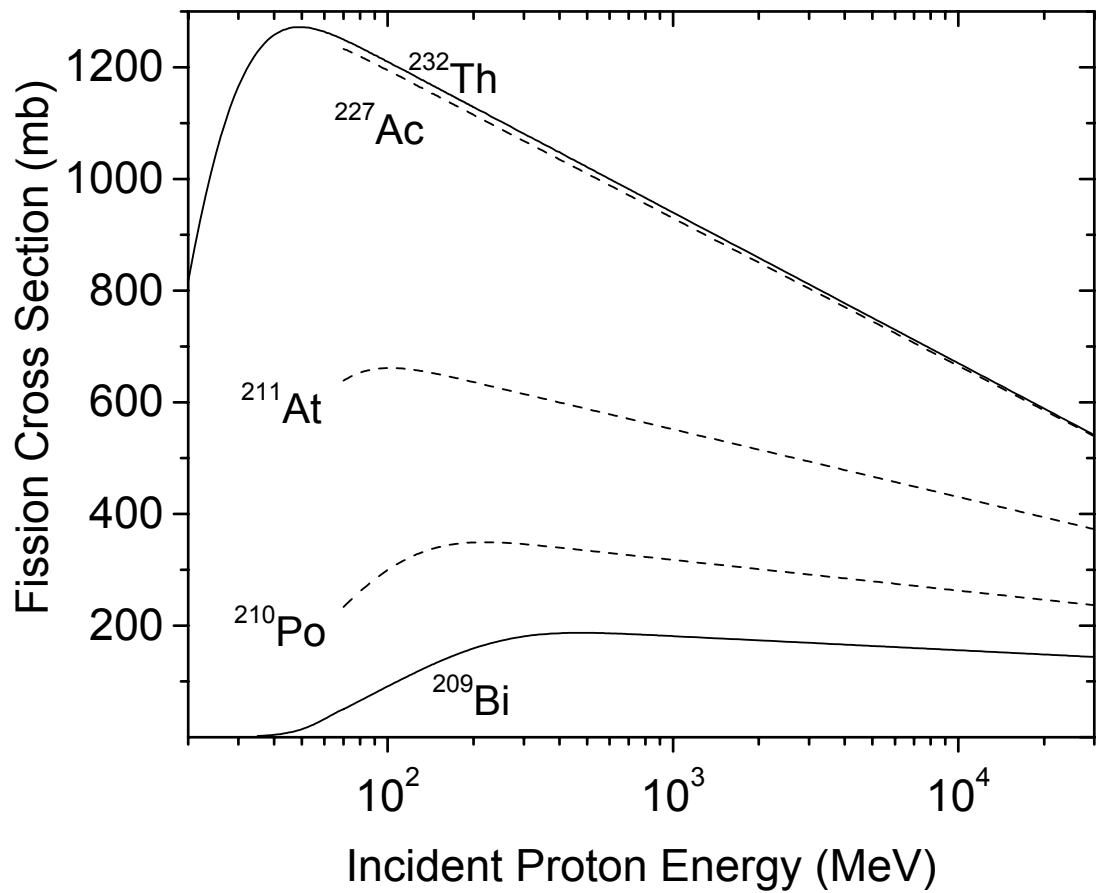

Fig. 3. Proton-induced fission cross sections for some nuclei with charge $Z = 83…92$. The solid lines represent predictions of the systematics of Prokofiev[4] for $^{209}$Bi(p,f) and $^{232}$Th(p,f) cross sections, for which abundant experimental data exist. The dashed lines show predictions of the extended systematics developed in the present work, for proton-induced fission of $^{210}$Po, $^{211}$At, and $^{227}$Ac. No experimental data are available for these nuclei.
ACKNOWLEDGEMENTS

The study is supported in part by The Swedish Foundation for International Cooperation in Research and Higher Education (STINT) and by the U.S. Department of Energy under contract no. W-7405-ENG-36.